\documentclass[aps,prl,reprint]{revtex4-1}
\usepackage{blindtext}
\usepackage{amsmath}
\usepackage{amsfonts}
\usepackage{graphicx}
\usepackage[colorlinks=true, allcolors=blue]{hyperref}

\newcommand{\veck}{{\vec{k}}}
\newcommand{\vecr}{{\vec{r}}}
\newcommand{\Z}{{\mathbb{Z}}}

\newcommand{\beq}{\begin{equation}}
\newcommand{\eeq}{\end{equation}}
\newcommand{\beqn}{\begin{eqnarray}}
\newcommand{\eeqn}{\end{eqnarray}}

\newcommand{\ii}{\mathrm{i}}

\newcommand{\U}{\mathrm{U}}

\begin{document}
\title{A no-go result for implementing chiral symmetries by locality-preserving unitaries in a $3$-dimensional Hamiltonian lattice model of fermions}
\author{Lukasz Fidkowski}
\affiliation{University of Washington, Seattle}
\author{Cenke Xu}
\affiliation{University of California at Santa Barbara}

\begin{abstract}

We argue that the chiral $\U(1)_A$ symmetry of a Weyl fermion cannot be implemented by a shallow depth quantum circuit operation in a fermionic lattice Hamiltonian model with finite dimensional onsite Hilbert spaces.  We also extend this result to discrete $\Z_{2N}$ subgroups of $\U(1)_A$, in which case we show that for $N_f$ Weyl fermions of the same helicity, this group action cannot be implemented with shallow depth circuits when $N_f$ is not an integer multiple of $2N$.

\end{abstract}

\maketitle








\section{Introduction}

Understanding how chiral symmetry of Weyl fermions can be implemented on the lattice is an important problem to multiple disciplines of physics. In the field of condensed matter theory, the well-known 't Hooft anomaly of the chiral $\U(1)_A$ symmetry is an archetypal example of the boundary of a symmetry protected topological (SPT) phase, which should preclude the possibility of realizing the $\U(1)_A$ as an exact on-site symmetry on the lattice. For the high energy theory community, realizing chiral fermions on the lattice is of both fundamental and practical importance, as the Standard Model and various version of Grand Unified Theories are chiral gauge theories. An celebrated no-go theorem by Nielsen and Ninomiya \cite{nielsen1981} explicitly concluded that the $\U(1)_A$ symmetry cannot be engineered in a local lattice model in three spatial dimensions for free fermions. Over the years this problem has been approached in several ways, for example by introducing an extra dimension \cite{kaplan1992}, so that the chiral fermions and the $\U(1)_A$ symmetry are realized at the boundary of a higher dimensional system, consistent with the general spirit of realizing the anomalous chiral symmetries at the boundary of a bulk SPT phase. It was also realized that the axial rotation symmetry could be an emergent symmetry in the infrared, which is implemented as a nonlocal transformation on a four-dimensional Euclidean spacetime lattice and action~\cite{luscher1998, luscher2000}.~\footnote{The feasibility of this approach in the Hamiltonian formalism, including the charge quantization of the axial rotation symmetry remains uncertain.}. 

In this work we will approach this problem from the standpoint of the fermionic analogue of the many-qubit model. That is, we will assume a many-body Hilbert space that is a graded tensor product of finite dimensional onsite fermionic Hilbert spaces, and investigate the action of chiral symmetry from a quantum information theoretic local unitary perspective.  This perspective has proven useful in elucidating aspects of anomalies in other contexts \cite{ChenWenGu,FidkowskiTSC,SWP,Metlitski_2015}.  Specifically, we ask the following question.  Suppose that a local Hamiltonian $H$ acting in such a Hilbert space realizes in its low energy limit a single Weyl fermion\footnote{Note that this is possible, for example by realizing two Weyl fermions in particle-number conserving band structure, and then gapping out one of the Weyl fermions with a (particle-number violating) Majorana mass term.}.  Is it then possible to find an operator $Q$ which is a sum of quasi-local terms such that $\exp\left(2\pi \ii Q\right)=1$, $[Q,H]=0$, and $Q$ 
coincides with the usual $\U(1)$ particle number associated to the Weyl fermion at low energies?  That is, can the chiral $\U(1)$ symmetry be implemented by a possibly non-onsite, but still quasi-local operator acting on this Hilbert space?

We argue that the answer to this question is no, for any number $N_f$ of Weyl fermions of the same helicity.  In particular, the $\U(1)_A$ axial symmetry of a Dirac fermion cannot be realized by a non-onsite operator of this form\footnote{The $\U(1)_A$ symmetry of a Dirac fermion can be viewed as the ordinary $\U(1)$ symmetry of a pair of Weyl fermions of the same helicity.}. We note that in Ref. \cite{creutz2002} a non-onsite free fermion operator $Q$ implementing axial $\U(1)_A$ symmetry in such a setting was written down; however, this $Q$ fails to satisfy $\exp\left(2\pi \ii Q\right)=1$, i.e. it does not generate a compact $\U(1)$ group.  There is in fact a simple topological argument which shows that any such free-fermion $Q$ must have a vanishing eigenvalue at some point in the Brillouin zone, which implies that in the thermodynamic limit excitations with arbitrarily small non-zero charges exist, a violation of charge quantization.  In the present work, we show that, more generally, an obstruction exists even at the level of interacting but still quasi-local $Q$.  We also show that the conclusion still holds if only a $\Z_{2N} \subset \U(1)_A$ subgroup is preserved, assuming that $N_f$ is not an integer multiple of $N$.

Our result can be viewed as a partial interacting generalization of the Nielsen and Ninomiya theorem \cite{nielsen1981}. Our argument relies on certain physical assumptions, the most important one being that it is impossible to have a chiral edge for a $(2+1)d$ commuting projector Hamiltonian.  The key part of the argument for the $\U(1)$ case is that a $\U(1)$ vortex line in the Majorana mass term carries $1d$ chiral modes, and a sufficiently local $Q$ would allow one to build a shallow depth circuit that inserts such a vortex at the boundary of a $2d$ membrane.  This circuit effectively builds a $2d$ chiral phase on this membrane from a trivial state, in particular giving it a commuting projector Hamiltonian, which is a contradiction.

\section{Assumptions and statement of the result}

We assume the Hilbert space is a graded tensor product, over the sites of some lattice, of $2$-dimensional fermionic Hilbert spaces with one even and one odd dimensional subspace.  We assume the existence of a fixed microscopic length scale $l$, that controls the locality of various operators as described below.  We will measure $l$ and all our other length scales in units of the lattice constant.  We assume the existence of a quasi-local Hermitian operator $Q=\sum_j Q_j$ with $\exp(2\pi \ii Q)=1$ and $Q_j$ acting near lattice site $j$.  We will not formalize the notion of quasi-local beyond the following.  First, $Q$ must satisfy the locality conditions of the Lieb-Robinson theorem \cite{hastingsLSM,hastingsNotes}, so that the unitary operators $\exp(\ii \tau Q)$, $0\leq \tau < 2\pi$ have a common Lieb-Robinson length \cite{hastingsLSM} bounded by $l$.  We will further consider operators of the form $\exp \left(\ii \tau \sum_j f_j Q_j\right)$, where $0\leq f_j\leq 1$ is a slowly varying function on the lattice.  We assume that these operators also have Lieb-Robinson length bounded by $l$.  Finally, we require that for any local operator $A_k$ supported on sites within distance $d$ of $k$, the operator norm of the difference between $\exp \left( \ii \tau \sum_j f_j Q_j\right) A_k \exp \left(- \ii \tau \sum_j f_j Q_j\right)$ and $\exp\left(\ii \tau Q\right) A_k \exp(- \ii \tau Q)$ is bounded by $Cd |\nabla f|\cdot||A_k||$ where $C$ is some constant, and $|\nabla f|$ the maximum gradient of $f$, in lattice units.  This condition just formalizes the idea that the action of a slowly spatially varying $\U(1)$ rotation on a local operator can be approximated by a constant $\U(1)$ rotation acting on that local operator, with the quality of the approximation controlled by the gradient of the spatial variation.

We will assume the existence of a local Hamiltonian $H$, possibly interacting, with the locality of the terms bounded by $l$, whose effective low energy theory is described by a Weyl fermion
\begin{align}\label{eq:HWeyl}
    H_{\rm{Weyl}}=\hbar v \sum_{|\veck|<\Lambda} \psi^\dagger (\veck)^T \left(\sum_{i=1}^3 k_i \sigma_i \right)\psi(\veck),
\end{align}
valid below a cutoff scale of $\hbar v / l$.  This means that the field operator
\begin{align*}
    \psi(\vecr)=N\sum_{|\veck|<\Lambda} e^{ \ii \veck\cdot\vecr} \psi(\veck)
\end{align*}
is quasi-local in the sense of being well approximated by an operator acting on a region of diameter $l$ around $\vecr$.  Furthermore, we will assume that the only low energy excitations are those of $H_{\rm{Weyl}}$, even on spatial manifolds of non-trivial topology.  This rules out the possibility of $H$ allowing additional gapped TQFT excitations, and means in particular that when $H$ is gapped out with a Majorana mass term below, the result is an invertible state.

Finally, we will assume that $[Q,H]=0$ and at low energies $Q$ generates the chiral $\U(1)$ particle number symmetry of $H_{\rm{Weyl}}$.  We will show that, taken together, these assumptions lead to a contradiction.

\section{No quasi-local $Q$ generating chiral $U(1)$}

We first compactify along one of the spatial directions, which we call $z$, and impose periodic boundary conditions along that direction, with length $l_z \gg l$.  We will parameterize the $z$ direction with a coordinate $z\in[0,l_z]$, with $0$ and $l_z$ identified.

\begin{figure}[tb]
\begin{center}
\includegraphics[width=0.9\linewidth]{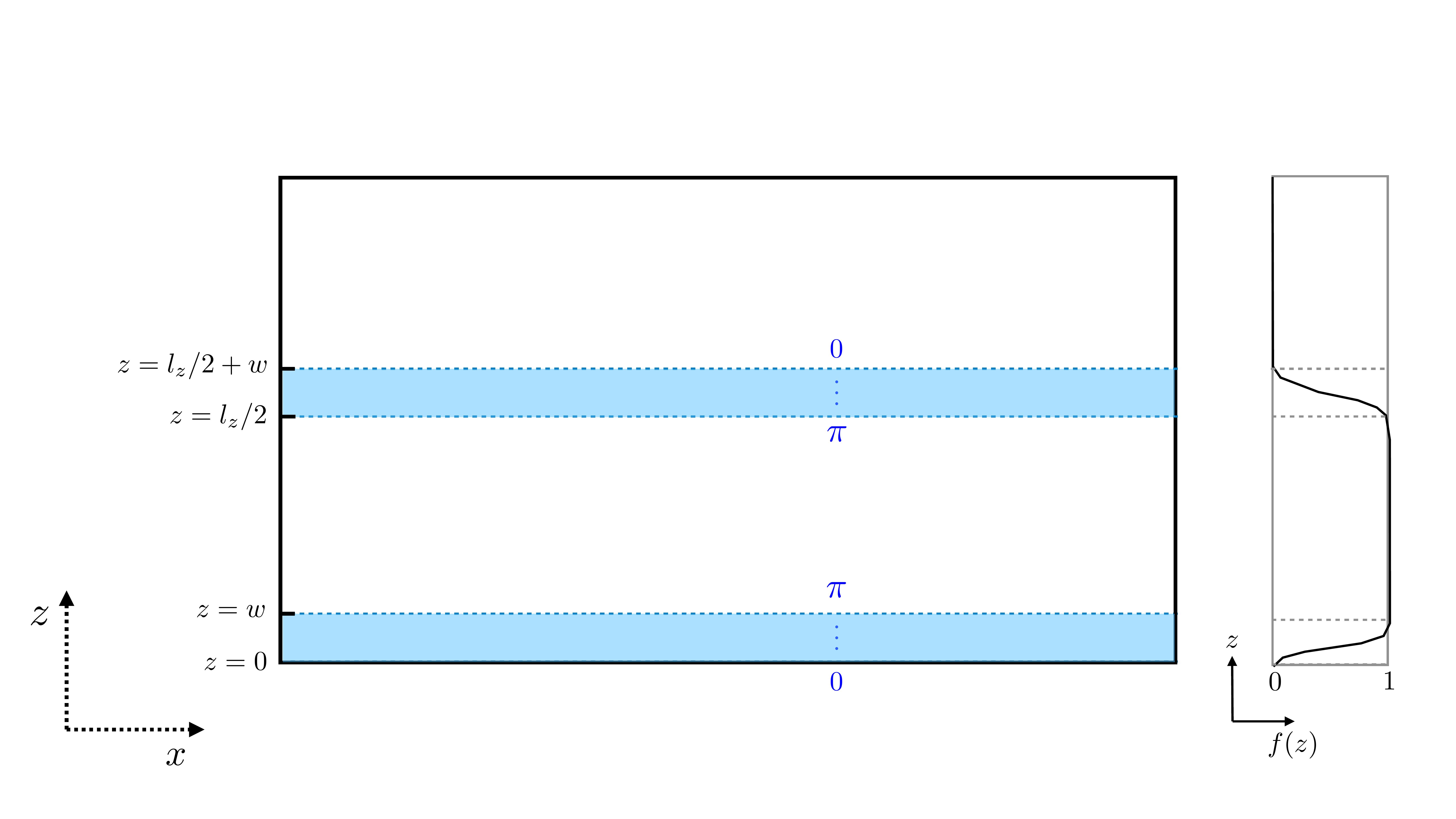}
\end{center}
\caption{Dimensionally reduced system, extended in the $x$ direction and $y$ direction (into the page).  The operator $V$ interpolates smoothly between a trivial operator for $l_z/2+w<z<l_z$ and a $\pi$ $\U(1)$ rotation for $w<z<l_z$ (acting as fermion parity $(-1)^F$).  The function $f(z)$ used to define $Q_f$ in Eq. \ref{eq:Qf} shown on the right.}
\label{fig:fig0}
\end{figure}

We now construct a unitary operator $V$ which smoothly interpolates between doing a $\pi$ $\U(1)$ rotation on half of the space (in the $z$ coordinate) and doing nothing on the other half, as illustrated in \ref{fig:fig0}.  Note that this is a $\pi$ rotation on the fermions, so a $2\pi$ rotation of the order parameter $\Delta(z)$.  To define $V$, we first pick $w$ such that $l \ll w \ll l_z$, and let $f(z)$ be an indicator function for $[0,l_z/2]$ smoothed out on the scale $w$; its defining properties are that it is equal to $1$ for $w < z < l_z/2$, it is equal to $0$ for $z$ outside of $[0,l_z/2+w]$, and it interpolates smoothly between those values in the intervening regions, with a maximum gradient of order $w^{-1}\ll l^{-1}$.  We then define

\begin{align} \label{eq:Qf}
Q_f = \sum_i f(z_i) Q_i
\end{align}
where for each $i$, $z_i$ is a point in the support of $Q_i$, and define $V=\exp\left(\ii \pi Q_f\right)$.

$V$ can be viewed as a finite time evolution of a bounded local pseudo-Hamiltonian, and has Lieb-Robinson length bounded by $l$.  Hence, after possibly increasing $w$ by an amount of order $l$, we can see that ${\tilde{V}}$ acts as the fermion parity operation $(-1)^F$ on all operators supported on the region $w<z<l_z/2$, and as the identity on all operators supported on the region $l_z/2+w < z < l_z$.  

We now perturb the Hamiltonian in our dimensionally reduced geometry as follows. We first define

\begin{align*}
H_{\Delta} = H + \int d\vecr \,\Delta \psi_1(\vecr)\psi_2(\vecr) + {\rm{h.c.}}
\end{align*}
We choose $|\Delta|$ sufficiently small so that the effect of the perturbation can be analyzed within the low energy free fermion theory, where it is just a Majorana mass term that gaps out the Weyl fermion Hamiltonian of eq. \ref{eq:HWeyl}. Note that this Majorana mass term induces a correlation length $\xi \sim \Delta^{-1}$, which we can assume to satisfy $l \ll \xi \ll w$.  As mentioned above, this gapped state must be invertible.  Using the fact that the conjectured classification of $(3+1)d$ invertible fermionic states is trivial, or by stacking with the conjugate invertible state, we can assume that this state is trivial, i.e. connected by a shallow depth circuit to a product state.  This means, in particular, that it has another parent Hamiltonian $H'_{\Delta}$ which is of local commuting projector form (obtained by conjugating the trivial commuting projector for the product state by this circuit).

Now, for $x<0$ we simply define $H_{x<0} = H_{\Delta}$.  We also define $H'_{x<0}=H'_{\Delta}$ to be the commuting projector parent Hamiltonian with the same ground state.  For $x>0$, we define $H_{x>0}$ as follows.  For $z$ away from the interval $[l_z/2,l_z/2+w]$, we take the local terms in $H_{x>0}$ to be identical to those of $H_{\Delta}$.  For $z$ in the interval $[l_z/2-\epsilon,l_z/2+w+\epsilon]$, we take the local terms to be those of $V H_{\Delta} V^{-1}$, where $\epsilon$ is some length scale with $l\ll \epsilon \ll l_z$.  This is a thickening of the interval $[l_z/2,l_z/2+w]$, but since $V$ acts as either the identity or fermion parity in the thickened regions, and all terms in $H_{\Delta}$ are fermion parity even, and hence not affected.  Carrying out the same procedure with $H'_{\Delta}$ yields a commuting projector Hamiltonian with the same ground state.

Let us now analyze $H_{x>0}$ in the low energy field theory.  Away from the range $l_z/2 < z < l_z/2+w$ it is just $H_{\Delta}$, while for $z$ in this range its terms are given by conjugating those of $H_{\Delta}$ by the slowly spatially varying $\U(1)$ rotation $v$.  Hence, these terms may be approximated by the action of a uniform $\U(1)$ rotation.  More precisely, for any such local operator $A$ of diameter $d$ localized at some $z$ coordinate $z_A$, we have by assumption that 

\begin{align}\label{eq:actiononA}
||{\tilde{V}} A {\tilde{V}}^{-1} - \exp\left(\ii f(z_A) Q\right) A \exp\left(-\ii f(z_A) Q\right)|| < \frac{Cd}{w} ||A||
\end{align}
Hence
\begin{align*}
    H_{x>0} &= H_{\rm{Weyl}} + H_{\rm{twisted}} + O(1/w)
\end{align*}
where
\begin{align*}
    H_{\rm{twisted}} = \int d\vecr \,\Delta(z)\psi_1(\vecr)\psi_2(\vecr) + {\rm{h.c.}}
\end{align*}
with $\Delta(z)$ being a complex number of amplitude $\Delta$ whose phase winds by $2 \pi$ as for $z \in [l_z/2,l_z/2+w]$.  Since the ground states of both $H_{x>0}$ and $H_{\rm{Weyl}} + H_{\rm{twisted}}$ are gapped, for sufficiently large $w$ the extra term of order $\frac{1}{w}$ above will have no effect, so that $H_{x>0}$ can be continuously connected to $H_{\rm{Weyl}} + H_{\rm{twisted}}$ without closing the gap.

But the situation with $H_{\rm{Weyl}} + H_{\rm{Majorana}}$ on one side of the interface and $H_{\rm{Weyl}} + H_{\rm{twisted}}$ is precisely a fundamental $2\pi$ vortex in the order parameter of the Majorana mass term, located at the effectively one dimensional interface $x=0$, as illustrated in figure \ref{fig:fig1}.  As derived in \ref{sec:vortex} below, a fundamental vortex hosts a gapless chiral mode with chiral central charge $1/2$.  Since this is a topological property, the same is true of the interface between $H_{x<0}$ and $H_{x>0}$, and also $H'_{x<0}$ and $H'_{x>0}$.  We thus have a situation where an interface between two $2d$ local commuting projector Hamiltonians hosts a gapless chiral mode.  If we fold the system at the interface, stacking the two gapped phases on top of each other, this yields a local commuting projector Hamiltonian with a chiral edge mode, which is believed to be impossible \cite{Kitaev_honeycomb}.  Hence we have derived a contradiction.

\begin{figure}[tb]
\begin{center}
\includegraphics[width=0.8\linewidth]{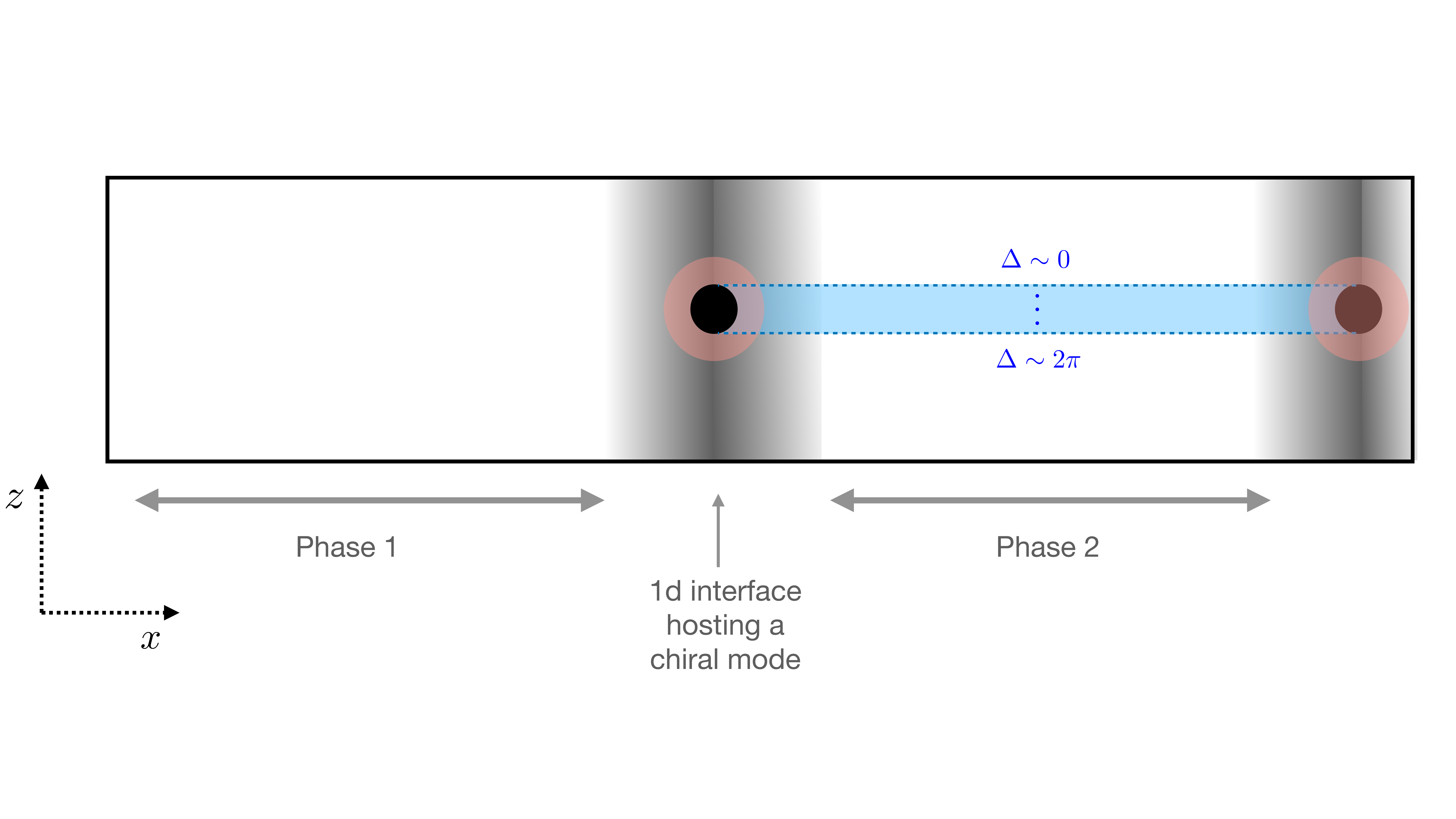}
\end{center}
\caption{A $1d$ spatial interface along the $y$ direction (into the page) between two different $2d$ gapped phases in the $x,y$ plane, at $x>0$ and $x<0$ respectively.  A $2\pi$ vortex in the order parameter $\Delta$ extends along this interface, as can be seen by the fact that the phase of $\Delta$ winds by $2\pi$ when crossing the blue region.}
\label{fig:fig1}
\end{figure}

\section{Generalization to discrete subgroups $\Z_{2N} \subset \U(1)$}

\begin{figure}[tb]
\begin{center}
\includegraphics[width=0.9\linewidth]{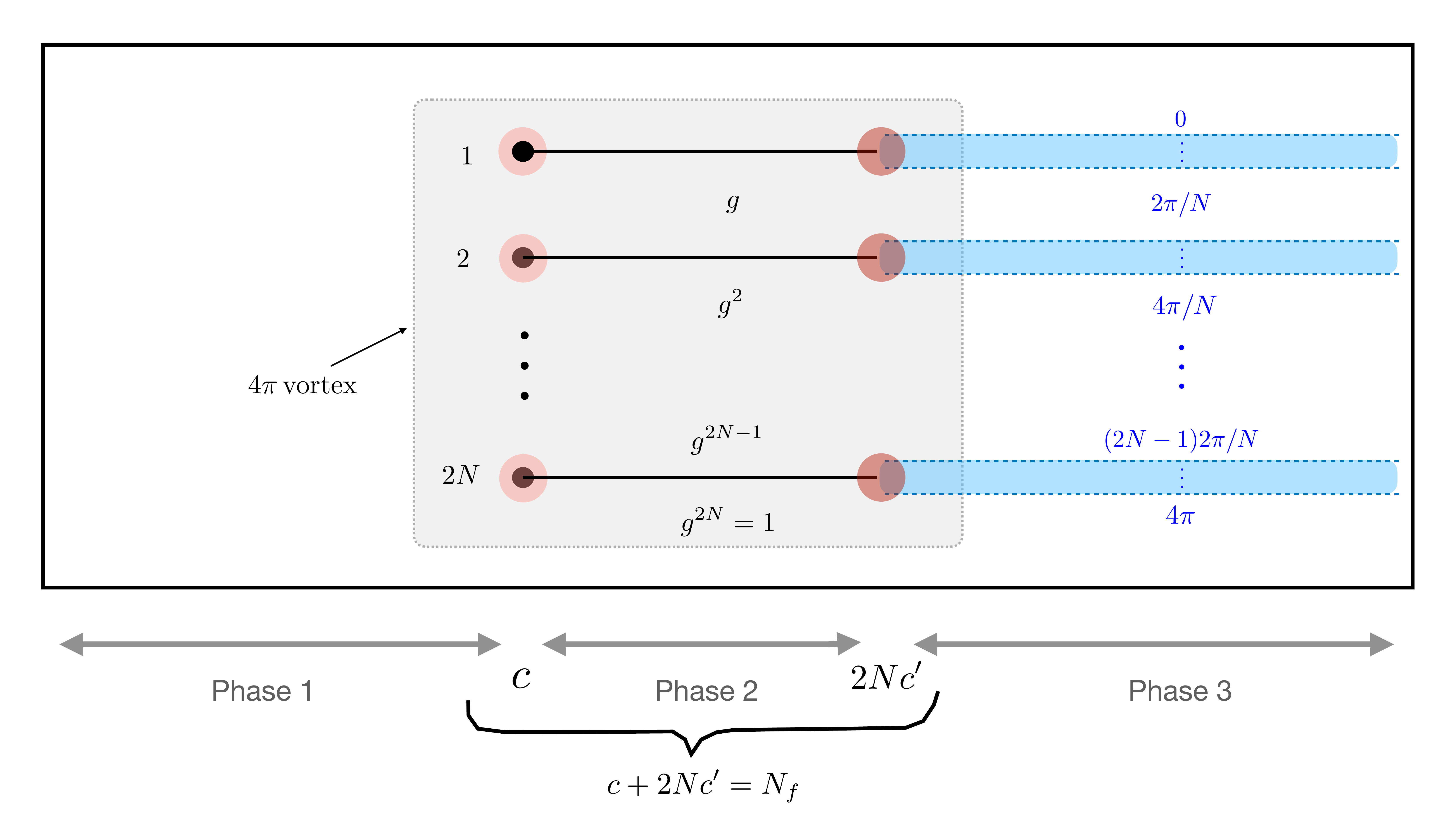}
\end{center}
\caption{An effective $4\pi$ vortex in $\Delta$ constructed from fusing $2N$ defects of a fundamental vortex of $\Z_{2N}\subset \U(1)$ symmetry.  Dimensionally reduced phases $1$ and $3$ can be described within the low energy field theory of $N_f$ Weyl fermions with Majorana mass term.}
\label{fig:fig2}
\end{figure}

Consider a theory of $N_f \neq 0 \mod 2N$ flavors of same helicity Weyl fermions, but instead of the full $\U(1)$ symmetry suppose that only a $\Z_{2N}$ subgroup is preserved.  We will show that in this case the $\Z_{2N}$ action cannot be realized by shallow depth quantum circuits in a graded tensor product Hilbert space.  The argument is again by contradiction, and is illustrated in figure \ref{fig:fig2}.

Again the geometry is dimensionally reduced to $2d$ along the $z$ direction.  We now stack $2N$ defects of the generator $g$ of the $Z_{2N}$ symmetry; as illustrated in figure \ref{fig:fig2} the state to the left of the defects is the original ground state (labeled `phase $1$'), whereas the state to the right of the defects (labeled `phase $2$') can be obtained from the ground state by multiplying by the truncated unitary actions of $g, g^2,\ldots, g^{2N}=1$ in the indicated regions (this is where we use the shallow circuit assumption, as shallow circuits can be truncated).  As noted in the $\U(1)$ case, phase $1$ is trivial and hence has a commuting projector parent Hamiltonian, and by essentially the same argument in the $\U(1)$ case, phase $2$ is trivial and has a commuting projector Hamiltonian as well.

Now, as we cross the branch cut surfaces emanating from each defect, the phase of $\Delta$ abruptly advances by $2\pi/N$.  This is a discontinuity in $\Delta$ and cannot be studied in the low energy theory of the Weyl fermion and Majorana mass term; however, we can imagine instead thickening the branch cut surfaces and advancing the phase slowly by $2\pi/N$ across these thickened surfaces.  This is labeled `phase 3' in figure \ref{fig:fig2} and can be studied using the low energy field theory.  The $1d$ interface between an abrupt branch cut (in phase $2$) and a slow one one (in phase $3$) is not necessarily gapped, and may carry some chiral central charge $c'$.  However, since the state immediately above and below the branch cuts can be disentangled into a product state, each such interface is effectively an interface between $2$-dimensional fermionic invertible gapped states, and hence $c'$ is an integer multiple of $1/2$.  Since the $2N$ different branch cuts are all related by symmetry, the value of $c'$ is the same for all.

Now we simply note that the grey shaded region encompassing phase 2 in figure \ref{fig:fig2} is effectively a $4\pi$ vortex in the Majorana mass term, when viewed from far away.  This means that its total chirality must be $N_f$, again by the argument in appendix.  This chirality is a sum of the contributions of the interfaces between phases $2$ and $3$, which contribute $2Nc'$, and some contribution $c$ from the defects (note that $c$ is {\emph{not}} necessarily a half integer multiple of $2N$).  Hence $c+2Nc'=N_f$, so $c=N_f-2Nc'$.  Thus if $N_f$ is not an integer multiple of $N$, then $c$ must be non-zero modulo $N$, and hence non-zero.  This is a contradiction for the same reason as in the $\U(1)$ proof, namely the fact that one cannot have a chiral edge mode between two $2d$ phases described by commuting projector Hamiltonians (phases $1$ and $2$ in figure \ref{fig:fig2}).

In this section we considered the situation where there is a discrete $\Z_{2N}$ chiral symmetry rather than a $\U(1)_A$ symmetry. As long as $N \geq 2$, this chiral symmetry still prohibits any fermion bilinear mass operator in the Weyl fermion Hamiltonian, though some higher order fermion operators are allowed. These higher order terms are obviously perturbatively irrelevant under RG flow, hence there will still be a $\U(1)_A$ emergent symmetry in the infrared. In the past few years, it has been gradually recognized that when the $\U(1)_A$ symmetry is broken down to its discrete subgroups on the lattice scale, Weyl fermions with emergent IR $\U(1)_A$ symmetry can be regularized as essentially a three-dimensional spatial lattice model, with the proper combination of short range interaction, flavor number, and flavor symmetries~\cite{wen2013,wangwen2013,youxu2014,youxu2014b,wangwen2020,Tong2021,Tong2022}. These results were obtained from various ways of demonstrating the absence of the 't Hooft anomaly of the discrete axial symmetries. It is worth noting that the absence of any 't Hooft anomaly is a stronger result that ours, which states that the symmetry cannot be implemented through a finite depth quantum circuit. For example, if $N=2$, i.e. if we have a $\Z_4$ subgroup of $\U(1)_A$, then the symmetry cannot be realized by finite depth circuits for any odd $N_f$, while it has been shown that the $\Z_4$ axial symmetry is completely anomaly free when $N_f = 0\,\mod\,16$~\cite{spinZ4,Tong2021,simon, Guo_2020, hsieh2018discrete}, and hence the symmetry should be realizable onsite in that case.  We note that if the symmetry can be realized by a locality preserving unitary \cite{Haah_2022} in the case $N_f=1$, then a standard argument would show that for any even $N_f$ the symmetry should be realizable by a shallow circuit\footnote{The argument is as follows: if a locality preserving unitary $X$ squares to a circuit, then $X \otimes X = (X\otimes 1) S (X\otimes 1) S$, where $S$ is the shallow depth swap operation.  But this is just $\left((X\otimes 1) S (X \otimes 1)^{-1}\right)(X\otimes 1)^2 S$, which is a product of shallow circuits, hence a shallow circuit.}

\section{Summary and Discussion}

In this work we showed that $\U(1)_A$ symmetry cannot be realized by a quasi-local charge $Q$, implying that the unitaries $\exp\left(\ii \tau Q\right)$ cannot be quasi-local shallow depth quantum circuits.  For the case of $\Z_{2N}$ and $N_f \neq 0\,\mod\, 2N$, we also showed that the unitaries generating the group action cannot be quasi-local shallow depth circuits.  In this case, there is still the possibility that these unitaries could be locality preserving but not shallow depth, i.e. they could be non-trivial quantum cellular automata (QCA); however, the fact that $U(1)$ is a continuous connected group makes this possibility unlikely \cite{Else2023}.  We note that while in-cohomology SPT phases always have a boundary action of symmetry by shallow-depth circuits \cite{Else_Nayak}, there exists a beyond cohomology SPT phase where the boundary action is a non-trivial QCA \cite{Fidkowski_Haah_Hastings2020}.  It would be interesting to relate the present work to this classification of boundary symmetry actions.

L.F. acknowledges useful conversations with David B. Kaplan, and support from NSF: DMR-1939864.  C.X. acknowledges the support from the Simons Foundation through the Simons Investigator program.

\appendix

\onecolumngrid

\section {Appendix: vortex line in a Majorana mass term condensate of a Weyl fermion hosts a $c=\frac{1}{2}$ chiral mode}\label{sec:vortex}

Consider a system in $3$ spatial dimensions which at low energy is described by a single Weyl fermion.  We will show that the core of a vortex line in the Majorana mass term hosts a $c=\frac{1}{2}$ chiral mode.  The Weyl fermion Hamiltonian is

\begin{align*}
    H_{\rm{Weyl}}=\hbar v \sum_{|\veck|<\Lambda} \psi^\dagger (\veck)^T \left(\sum_{i=1}^3 k_i \sigma_i \right)\psi(\veck)
\end{align*}
where $\Lambda$ is a UV cutoff and without loss of generality we work in finite volume.  Here $\psi$ has $2$ components, the dagger denotes a hermitian conjugate in Fock space, and the $T$ denotes the transpose of the $2$-component vector.  The $\sigma_i$, $i=1,2,3$, are the usual Pauli matrices, and $v$ is a velocity.  For convenience we will work below in units where $\hbar v=1$.  The local operators are

\begin{align*}
    \psi(\vecr)=N\sum_{|\veck|<\Lambda} e^{\ii \veck\cdot\vecr} \psi(\veck)
\end{align*}
where $N$ is a normalization constant.  We now introduce the Majorana mass term with a vortex

\begin{align*}
  H_{\rm{vortex}} = \int d\vecr \,\Delta(\vecr)\psi_1(\vecr)\psi_2(\vecr) + {\rm{h.c.}}
\end{align*}
where $\psi_1$ and $\psi_2$ are the two components of $\psi$.
For $\Delta(\vecr)$ we take a vortex line configuration in the $z$ direction.  That is, $\Delta(\vecr)$ is independent of $z$, is equal to $0$ at $x,y=0$ and is non-zero elsewhere, with $|\Delta(\vecr)|$ approaching a non-zero constant $\Delta$ away from $x,y=0$.  $\Delta(\vecr)$ has a phase that winds around by $2\pi$ around the unit circle in the $xy$ plane.  Our result depends only to the topology of the vortex configuration and is not sensitive to its precise form; we will specify a convenient form below.  Because the vortex configuration retains translational symmetry in the $z$ direction, it will be convenient to work with

\begin{align*}
  \psi(x,y;k_z)=N'\sum_{|k_x^2+k_y^2|<\Lambda} e^{\ii (k_x x+k_y y)}\, \psi(k_x,k_y,k_z)
\end{align*}
where $N'$ is another normalization constant.  The full Hamiltonian for the vortex is then

\begin{align*}
H&=H_{\rm{Weyl}}+H_{\rm{vortex}} \\ &= \sum_{k_z} \left[ \sum_{k_x,k_y} \psi^\dagger (\veck)\left(\sum_{i=1}^3 k_i \sigma_i\right)\psi(\veck) + \left(\int dx\,dy\, \Delta(x,y)\psi_1(x,y;k_z) \psi_2(x,y;-k_z)+{\rm{h.c.}}\right)\right]
\end{align*}
We now define operators
\begin{align}\label{eq:chi_def}
    \chi(\veck)=\ii \sigma_2\psi^\dagger(-\veck)
\end{align}
or, in components, $\chi_1(\veck)=\psi_2^\dagger(-\veck)$, $\chi_2(\veck)=-\psi_1^\dagger(-\veck)$.  Our many-body Hilbert space is the tensor product of $2$-level fermionic Hilbert spaces spanned by $\psi(\veck)$ for all $\veck$, but can now equivalently be viewed as the Hilbert space of $2$-level fermionic Hilbert spaces spanned by $\psi(\veck)$ and $\chi(\veck)$ for $\veck$ with $k_z>0$.\footnote{We ignore for the moment the zero mode $k_z=0$; this can be done e.g. by imposing anti-periodic boundary conditions in $z$.} Using $\psi^\dagger(-\veck)^T=\chi(\veck)^T(- \ii \sigma_2)^T$ and $\psi(-\veck) =- \ii \sigma_2\chi^\dagger(\veck)$, which follow from eq. (\ref{eq:chi_def}), $H_{\rm{Weyl}}$ can then be written as

\begin{align*}
    H_{\rm{Weyl}}&=\sum_{k_x,k_y} \left[\sum_{k_z>0} \psi^\dagger(\veck)^T \left(\sum_{i=1}^3 k_i \sigma_i \right)\psi(\veck) + \sum_{k_z<0}\psi^\dagger(\veck)^T \left(\sum_{i=1}^3 k_i \sigma_i \right)\psi(\veck)\right]\\ &= \sum_{k_x,k_y; k_z>0} \left[\psi^\dagger(\veck)^T \left(\sum_{i=1}^3 k_i \sigma_i \right)\psi(\veck) + \psi^\dagger(-\veck)^T \left(-\sum_{i=1}^3 k_i \sigma_i \right)\psi(-\veck)\right] \\ &=\sum_{k_x,k_y; k_z>0} \left[\psi^\dagger(\veck)^T \left(\sum_{i=1}^3 k_i \sigma_i \right)\psi(\veck) + \chi(\veck)^T (- \ii \sigma_2)^T \left(-\sum_{i=1}^3 k_i \sigma_i \right)(- \ii \sigma_2)\chi^\dagger(\veck)\right] \\&=\sum_{k_x,k_y; k_z>0} \left[\psi^\dagger(\veck)^T \left(\sum_{i=1}^3 k_i \sigma_i \right)\psi(\veck) + \chi(\veck)^T (- \ii \sigma_2)^T \left(-k_1\sigma_1+k_2\sigma_2-k_3\sigma_3 \right)^T( \ii \sigma_2)^T\chi^\dagger(\veck)\right]\\&=\sum_{k_x,k_y; k_z>0} \left[\psi^\dagger(\veck)^T \left(\sum_{i=1}^3 k_i \sigma_i \right)\psi(\veck) -  \chi^\dagger(\veck)^T (\ii \sigma_2) \left(-k_1\sigma_1+k_2\sigma_2-k_3\sigma_3 \right)(-\ii \sigma_2)\chi^\dagger(\veck)\right]\\&=\sum_{k_x,k_y; k_z>0} \left[\psi^\dagger(\veck)^T \left(\sum_{i=1}^3 k_i \sigma_i \right)\psi(\veck) -  \chi^\dagger(\veck)^T  \left(\sum_{i=1}^3 k_i \sigma_i \right)\chi^\dagger(\veck)\right]
\end{align*}
In the same way, the Majorana mass term can be written as

\begin{align*}
    H_{\rm{vortex}}=\sum_{k_z>0} \int dx\,dy\,\Delta(x,y)\left( \psi_1(x,y;k_z) \chi_1^\dagger(x,y;k_z)+\psi_2(x,y;k_z) \chi_2^\dagger(x,y;k_z)\right) + {\rm{h.c.}}
\end{align*}
Thus

\begin{align} \label{eq:H}
    H&=\sum_{k_z>0}\sum_{k_x,k_y} \left(\psi^\dagger(\veck)^T \left(\sum_{i=1}^3 k_i \sigma_i \right)\psi(\veck) -  \chi^\dagger(\veck)^T  \left(\sum_{i=1}^3 k_i \sigma_i \right)\chi^\dagger(\veck)\right) \\&- \sum_{k_z>0}\int dx\,dy\,\Delta(x,y) \chi^\dagger(x,y;k_z)^T \psi(x,y;k_z)+ {\rm{h.c.}}
\end{align}
This is just the $k_z>0$ part of the Hamiltonian describing a vortex line in the ordinary mass term of a Dirac fermion. 
Diagonalizing such a free fermion particle number conserving Hamiltonian is known to result in low energy theory consisting of a chiral mode bound to the vortex line.  That is, for each $k_z \ll \Lambda$, there is a unique low energy state $\eta(k_z)$, extended in the $z$ direction with wave vector $k_z$ but localized near $x,y=0$ in the $x,y$ directions, with energy eigenvalue $\hbar v k_z$, where $v$ is a velocity.  Now, in the case of our Hamiltonian (eq. \ref{eq:H}), only the $k_z>0$ part of the spectrum is physical.  This is precisely the spectrum of a $c=\frac{1}{2}$ chiral Majorana fermion mode $\eta(z)$ in $1+1$ dimensions, since $\eta(z) = \eta^\dagger(z)$ implies $\eta(-k_z)=\eta^\dagger(k_z)$.

This demonstrates the claim we set out to prove, but for completeness we now show that a vortex line in the mass term of a Dirac fermion hosts a chiral mode.  We begin with the same Hamiltonian as in eq. \ref{eq:H}, except without a restriction on $k_z$:

\begin{align} \label{eq:newH}
    \int d\vecr \left[\psi^\dagger(\vecr)^T\left(\ii \sum_{j=1}^3 \sigma_j \partial_j\right)\psi(\vecr)-\chi^\dagger(\vecr)^T\left(\ii \sum_{j=1}^3 \sigma_j \partial_j\right)\chi(\vecr)+m(x,y)\chi^\dagger(\vecr)^T \psi(\vecr)+{\rm{h.c.}}\right]
\end{align}
Here we have changed the notation from $\Delta(x,y)$ to $m(x,y)$ for what is now a mass term.  We will construct a vortex by first diagonalizing and finding the low energy spectrum of a domain wall, where $m(x,y)=m(y)$ interpolates between $-m$ at $y\rightarrow -\infty$ and $m$ at $y\rightarrow \infty$.  In this case translational symmetry in the $x$ and $z$ directions is retained, so the Hamiltonian can be written as

\begin{align*}
   H_{\rm{domain}\, \rm{wall}} &= \sum_{k_x,k_z} \int dy\,\psi^\dagger(k_x,k_z;y)^T\left(\ii \sigma_2 \partial_y+k_x\sigma_1+k_z\sigma_3\right)\psi(k_x,k_z;y)\\&-\chi^\dagger(k_x,k_z;y)^T\left(\ii \sigma_2 \partial_y+k_x\sigma_1+k_z\sigma_3\right)\chi(k_x,k_z;y) \\ &+m(y)\chi^\dagger(k_x,k_z;y)^T\psi(k_x,k_z;y)+{\rm{h.c.}}
\end{align*}
Focusing on eigenstates for a fixed $k_x,k_z$, the eigenvalue equation reads

\begin{align}\label{eq:1deig}
\left[\left(\ii\sigma_2\partial_y+k_x\sigma_1+k_z\sigma_3\right)\tau_3 + m(y)\tau_1\right]v(y) = E v(y).
\end{align}
Here we have combined $\psi$ and $\chi$ into a single $4$-component vector $v$.  The matrices $\sigma_j$ are now implicitly tensored with the identity on the $(\psi,\chi)$ space, and the matrices $\tau_j$ are Pauli matrices acting on this $(\psi,\chi)$ space, tensored with the identity on the original spin space.  The eigenvalue equation may be re-written as:

\begin{align*}
\partial_y v = \ii \sigma_2\tau_3\left[ \left(k_x \sigma_1+k_z\sigma_3\right)\tau_3 + m(y)\tau_1-E\right] v
\end{align*}
Let us first analyze the point $k_x=k_z=0$, in which case the eigenvalue equation reduces to:

\begin{align*}
  \partial_y v = \left(m(y)\sigma_2 \tau_2 - \ii \sigma_2\tau_3 E \right) v
\end{align*}
This is just two copies of the familiar Jackiw-Rebbi \cite{JR} soliton equation in $1+1$ dimension.  It is known that the only low energy solutions exist for $E=0$, in which case the equation reads $\partial_y v = m(y)\sigma_2 \tau_2$.  There are two solutions, both of which are eigenvectors of $\sigma_2 \tau_2$ with the same eigenvalue. The eigenvalue, $\pm 1$, is determined by whether $m(y)$ is positive or negative at large $y>0$; in our case it is $+1$.  The modes described by these two solutions are exponentially localized at $y=0$. 

For $k_x,k_z\neq 0$, we can simply treat the $(k_x\sigma_1+k_z\sigma_3)\tau_3$ term as a perturbation and work to first order in perturbation theory.  Note that $\sigma_1\tau_3$ and $\sigma_3\tau_3$ both commute with $\sigma_2 \tau_2$.  Within the two dimensional $+1$ eigenspace of $\sigma_2 \tau_2$, the two operators $\sigma_1\tau_3$ and $\sigma_3\tau_3$, together with $\tau_2$ have the commutation relations of the usual Pauli matrices; let us denote them by $\alpha_1, \alpha_3$, and $\alpha_2$ respectively.  The low energy theory on the domain wall $y=0$ is then given by a $2$ dimensional Dirac Hamiltonian $k_x \alpha_1+k_z\alpha_3$.  Now, to form a vortex in the original problem, we will introduce, within this low energy theory, a $1$d domain wall at $z=0$.  In order for this to be topologically equivalent to a vortex in the original problem, the mass term we introduce must correspond to the case of imaginary $m(x,y)$ in eq. \ref{eq:newH}, which translates to $\tau_2$ in the $4$-component notation, which in turn translates to $\alpha_2$ in the $2$-component notation in the low energy theory living at $y=0$.

Now, $m(z) \alpha_2$ (with $m(z)\rightarrow \pm m$ at $z\rightarrow \pm \infty$) is simply a domain wall in the usual mass term of the Dirac Hamiltonian $k_x \alpha_1+k_z\alpha_3$ (we are dropping an overall factor of $\hbar$ times a velocity), and it is well known that it hosts a single chiral mode at $z=0$.  One may show this by once again writing down the eigenvalue equation for low energy states, which is now a single copy of the Jackiw-Rebbi soliton for a $2$-component vector $w$:

\begin{align*}
    \partial_z w = (\ii k_x\alpha_2 + m(z)\alpha_1 - \ii E\alpha_3) w
\end{align*}
For $k_x=0$ there is an $E=0$ mode which is an eigenvector of $\alpha_1$.  For non-zero $k_x$ we may again use first order perturbation theory; the perturbation is simply $k_x\alpha_1$, so its eigenvalue is $k_x$.  A mode with such 
a dispersion is just a chiral mode, as was to be shown.

\bibliography{sample}

\end{document}